\documentclass{article}[11pt]
\usepackage{amsmath,amsfonts,amsthm,amssymb}
\usepackage[blocks]{authblk}
\usepackage{hyperref}
\usepackage[utf8]{inputenc}
\usepackage[T1]{fontenc}
\usepackage[margin={2cm,1cm}]{geometry}
\usepackage{cite}
\usepackage{authblk}
 \usepackage{nicematrix,tikz}
\date{}
\usepackage[utf8]{inputenc}
\usepackage[upright]{fourier}
\usepackage{tikz}
\usetikzlibrary{matrix}
\usepackage{fullpage,amsmath}

\begin{document}
\pagenumbering{gobble}
	\title{Relation between nonclassical features through logical qudits}

\author[1]{\underline{Sooryansh Asthana}\thanks{sooryansh.asthana@physics.iitd.ac.in}}
\author[2]{V. Ravishankar\thanks{vravi@physics.iitd.ac.in}}
\affil[1,2]{Department of Physics,
Indian Institute of Technology Delhi, 
New Delhi, India-110016.}

	\maketitle

\vspace{0.5cm}
Scalable modern-time fault-tolerant quantum computation and quantum communication in a network employ a large number of physical qubits. For example,  IBM is reported to have made a 127-qubit quantum computer.  Unlike classical computation, quantum computation employs different types of logical qubits and qudits in terms of physical multiqubit and multiqudit systems respectively.  Given this, of particular interest to us is to enquire on how quantum coherence in logical qubits is a manifestation of underlying quantum correlations in constituent physical multiqubit systems and vice-versa. In a recent work \cite{Asthana22}, we have shown that there is a reciprocity in
nonclassical correlations in physical multiqubit systems and coherence in a single logical qubit system. Subsequently, we have generalised the framework to higher dimensional quantum systems \cite{asthana2022a}. The crux of this study is that a single nonclassicality condition derived for quantum coherence in a logical system detects more than one type of nonclassicality in Hilbert spaces of nonidentical dimensions.
\vspace{1cm}
\section{Introduction}
Modern-day quantum computation and quantum communication in a network employ a large number of physical qubits, as in the example of superconducting qubits \cite{gambetta2017building, huang2020superconducting, jurcevic2021demonstration, lee2020quantum}. For example, IBM quantum computer is reported to have 127 physical qubits and IBM has promised a quantum computer consisting of 1000 physical qubits by 2023 \cite{cho2020ibm}. However, as noted in \cite{cho2020ibm}, the IBM quantum computers have only a handful of logical qubits. The demand of  fault-tolerance necessitates different types of logical qubits in terms of their physical constituent qubits \cite{steane1999efficient, barends2014superconducting}.   Similarly, noise-resilient communication protocols employ logical qubits which are composed of physical multiqubit systems. In fact, quantum mechanics allows for coherent logical qubits which involve coherent superpositions of underlying physical multiqubit systems.  On the other hand, classical logical bits do not involve any coherent superposition of constituent bits. So, one may expect that quantum coherence in a logical qubit or qudit may emerge from underlying quantum correlations in physical multiqubit systems and vice-versa.

Given this,  in this work, we enquire on how conditions for different nonclassical correlations in a physical multiqudit systems emerge from those for quantum coherence in a logical system and vice-versa. For this, we generalise the framework developed recently which is applicable only to logical qubits \cite{Asthana22, asthana2022a}.  We show that there is a reciprocity between conditions for nonclassical correlations in physical multiqudit systems and those for coherence in a logical qudit system. 

\section{Framework}
\label{Procedure}
It is convenient to start with the coherence condition for a logical qudit state expressed in the computational basis $\{|i\rangle_L\}$, where each logical qudit is a direct product of states of $N$ physical qudits, i.e., 
$|i\rangle_L \equiv |i\rangle^{\otimes N};~ 0\leq i \leq (d-1)$.  A condition for quantum coherence in the basis, $\{|i\rangle_L\}$ is given as,
\begin{align}
{\cal C}_L: \langle {\cal A}_L\rangle > c,~~~~~~~~~~~   ~c> 0,
\end{align}
 with the stipulation that it is maximally obeyed by the state
$|\psi\rangle_L \equiv\dfrac{1}{\sqrt{d}}\sum_{i=0}^{d-1}|i\rangle_L$.  The operator ${\cal A}_L$ has the following equivalent forms,
 \begin{align}
 {\cal A}_L=\sum_{i,j=0}^{d-1} a_{ij}^{(L)}|i\rangle_L{}_L\langle j|=\sum_{i, j=0}^{d-1}a^{(L)}_{ij}\Big(|i\rangle\langle j|\Big)^{\otimes N}.  
 \end{align}
The task is to infer the underlying quantum correlation in the physical multiqudit states. This would require identification of  distinct sets of operators acting over  physical qudits that give rise to the witness ${\cal C}_L$. We show that such sets do indeed exist and may be constructed in a manner that the underlying correlation is brought out. The operators in these sets may, in turn, be employed to construct conditions for nonclassical correlations in the underlying multiqudit system as follows:
\begin{enumerate}
    \item We resolve the operator ${\cal A}_L$ as a direct product of the operators $A_1, A_2, \cdots, A_N$, where $A_{\alpha}$ acts on the logical qudit labelled $\alpha$. The resolution is not unique. Consider, for example, the following prescription:\\
       $\big({\cal A}_L\big)_{ij} \neq 0 \implies\prod_{\alpha=1}^N\big(A_{\alpha}\big)_{ij} =\big({\cal A}_L\big)_{ij},$ where $\prod_{\alpha=1}^N\big(A_{\alpha}\big)_{ij} $ represents element-by-element multiplication.    Note that this prescription  allows for many choices of sets of operators $A_1, A_2, \cdots, A_N$ for the same logical operator ${\cal A}_L$. The action of the operator ${\cal A}_L$ on the state $|\phi_L\rangle$ is identical to that of the  tensor product of operators $A_1, A_2, \cdots, A_N$. 

       \item Let $\Big\{A_{\alpha}^{(k)}\Big\}$ represent the $k^{\rm th}$ resolution of ${\cal A}_L$. Then, the operator ${\cal A}_L$ can be variously expressed as $\sum_k c_k \Pi_{\alpha=1}^N A_{\alpha}^{(k)}$, where the sole condition on $c_k$ is that they are normalised weights adding upto one. The choices of the weights and the bound on the ensuing inequality depend on the notion of classicality under consideration. Of all the resolutions of ${\cal A}_L$, those resolutions are of particular interest for us in which the operator acting over the same qudit are noncommuting, i.e., $\Big[A_{\alpha}^{(k)}, A_{\alpha}^{(l)}\Big] \neq 0$. These resolutions bring out nonclassical correlations in the underlying physical system.
\end{enumerate}

\section{Example: Quantum correlation in a two-qutrit system from coherence for a single logical qutrit}
 Let the basis states in the Hilbert space of logical qutrits be $|0\rangle_L, |1\rangle_L, |2\rangle_L$. In this case, we choose ${\cal A}_L$ to be a generalised Pauli shift operator, i.e., ${\cal A}_L \equiv \sum_{i=0}^2|i+1\rangle_L{}_L\langle i|$, where the addition is modulo 3.
 We choose a condition for coherence $|\langle {\cal A}_L \rangle|> c, c\in [0, 1)$, with respect to the basis $\{|0\rangle_L, |1\rangle_L, |2\rangle_L\}$.
 The state $|\phi\rangle_L\equiv \frac{1}{\sqrt{3}}\Big(|0\rangle_L+|1\rangle_L+|2\rangle_L\Big)$ obeys this condition, as $\langle \phi_L|{\cal A}_L|\phi_L\rangle =1$.  We now move on to show how the condition for coherence, $|\langle {\cal A}_L\rangle|> c$, in a logical system, gives rise to a condition for quantum correlations in a two-qutrit system. 

Let each logical qutrit be composed of a pair of qutrits, i.e.,  $|0\rangle_L \equiv |00\rangle, |1\rangle_L\equiv |11\rangle, |2\rangle_L\equiv |22\rangle$. So, the state $|\phi\rangle_L$ assumes the form    $|\phi\rangle_L \equiv \dfrac{1}{\sqrt{3}}\Big(|00\rangle+|11\rangle+|22\rangle\Big)$.
 Following the procedure described in section (\ref{Procedure}), the logical operator, ${\cal A}_L$ can be mapped to any one of the following  physical qutrit operators,
\begin{align}
{\cal A}_L\to &\begin{cases}
      X_1X_2,\\
      (X_1Z_1)(X_2Z_2^2),\\
       (X_1Z_1^2)(X_2Z_2).
    \end{cases} 
\end{align}
 The operators $X_1$ and $X_2$ of the first and the second qutrit  have the same forms in the computational bases as that of ${\cal A}_L$ in the basis $\{|0\rangle_L, |1\rangle_L, |2\rangle_L\}$.  The operator $Z$ is defined as: $Z\equiv {\rm diag}~(1, \omega, \omega^2)$, where $\omega$ is the cube root of identity. Obviously, $\omega^3 =1$  and  $[X_1, Z_1]\neq 0$, $[X_2, Z^2_2]\neq 0.$
\subsection{Entanglement in a two-qutrit system}
  If we choose a pair of operators, $X_1X_2, ~X_1Z_1X_2Z_2^2$, the condition for coherence, $|\langle {\cal A}_L\rangle|> c$, yields the following condition: 
\begin{align}
|\langle X_1X_2+X_1Z_1X_2Z_2^2\rangle| >2c,
\end{align}
which is indeed the condition for entanglement in two-qutrit systems if $ \frac{1}{2}\leq c <1$. In fact, as we decrease the value of  $c$ in the interval $\Big[\frac{1}{2}, 1\Big)$, the coherence witness for logical qutrit and hence, entanglement witness for the underlying two-qutrit systems becomes more encompassing. 

\subsection{Nonclassical correlation in two different bases in a two-qutrit system}
The condition for coherence $|\langle {\cal A}_L\rangle| >c$, also yields the following two conditions, $|\langle X_1X_2\rangle|> c$ and $|\langle X_1Z_2^2X_2Z_2\rangle|>c$ for all nonzero values of $c$. If these two conditions are simultaneously satisfied, they detect states having nonzero  correlations in the following two eigenbases of locally noncommuting operators: (i) the first basis  consists of eigenvectors $X_1$ and  $X_2$, and, (ii) the second basis consists of eigenvectors of  $X_1 Z_1$ and $X_2Z_2^2$. In this manner, depending upon the choices of weights and bound on the ensuing inequalities, the condition for coherence $|\langle {\cal A}_L\rangle| >c$ leads  to conditions for different nonclassical correlations in the underlying two-qutrit system.

\section{Conclusion}
In summary, we have laid down a formalism employing logical qudits to study the reciprocity between nonclassical correlations in physical multiqudit systems and quantum coherence in a single logical qudit. As an example, we have shown how condition for quantum coherence in a logical qutrit gives rise to condition for quantum entanglement in a two-qutrit system. In this way, this work shows interrelation of different nonclassical features. 


\end{document}